\documentclass[twocolumn,prb]{revtex4}
\usepackage{graphicx}
\begin{document}
\draft
\preprint{to be published to Physical Review B, 1 Oct. (2002).}
\title{Hall coefficient of La$_{1.88-y}$Y$_y$Sr$_{0.12}$CuO$_4$ ($y=0, 0.04$) at low temperatures under high magnetic fields}
\author{T. Suzuki, and T. Goto}
\address{Department of Physics, Sophia University, 7-1 Kioi-cho, Chiyoda-ku Tokyo 102-8554, Japan}
\author{K. Chiba, M. Minami, Y. Oshima, and T. Fukase}
\address{Institute for Materials Research, Tohoku University, Sendai 980-8577, Japan}
\author{M. Fujita, and K. Yamada}
\address{Institute for Chemical Research, Kyoto University, Uji Kyoto, 610-0011, Japan}
\date{\today}
\begin{abstract}
The Hall coefficient in the low-temperature tetragonal phase and the mid-temperature orthorhombic phase of La$_{1.88-y}$Y$_y$Sr$_{0.12}$CuO$_4$ ($y=0, 0.04$) single crystals is measured under high magnetic fields up to 9 T in order to investigate the detailed behavior of the transport properties at low temperatures in the stripe phase.
When the superconductivity is suppressed by high magnetic fields, the Hall coefficient has negative values in low temperatures, and the temperature region of the negative values spreads as increasing magnetic fields.
This result indicates that the Hall coefficient in the stripe phase around $x=0.12$ is a finite negative value, not zero.
\end{abstract}
\maketitle
\indent
Curious properties associated with "1/8-probrem"  have been investigated over the last decade in the world and yet this is one of the centre of wide interest in high-{\it T}$_{\rm c}$ superconductors.
Let us begin this paper by reviewing something rather old.
La$_{2-x}$Ba$_x$CuO$_4$ (LBCO) and La$_{2-x}$Sr$_x$CuO$_4$ (LSCO) show the local minimum of {\it T}$_{\rm c}$ around $x\sim 0.12$~\cite{Tc-dip1,Tc-dip2}.
It has been predicted that the disappearance in LBCO or suppression in LSCO of the superconductivity is correlated with the structural change : the structural phase transition from the mid-temperature orthorhombic phase (OMT, the space group Bmab) to the low-temperature tetragonal phase (TLT, the space group P4$_2$/ncm) in LBCO and the precursor of the transition in LSCO around $x=0.12$~\cite{LTT,Fukase}.
This prediction is verified by analysis of the crystal structure in Rare-earth doped LSCO system, which undergoes the structural phase transition to the TLT phase~\cite{Buchner,Crawford,Suzuki1}.
In addition to the disappearance or the suppression of the superconductivity, the two related phenomena appear in these system.
One is magnetic order~\cite{Kumagai,Goto}, and the second is anomalous changes of transport properties, such as the Hall coefficient and the thermoelectric power~\cite{Sera,Uchida}.
It is suggested that the suppression of the superconductivity around $x=0.12$ will be related with not only the crystal structure but also changes in magnetic and electronic states.
\\
\indent
The recent development on this problem is followed from the "stripe model"~\cite{Tranq1,Tranq2,Tranq3}.
According to this model, the long-range modulated charge and spin ordering is stabilized in the TLT phase.
Afterwards, magnetic superlattice peaks are observed by neutron diffraction in the orthorhombic superconducting phase La$_{1.88}$Sr$_{0.12}$CuO$_4$ single crystal, while no peaks associated with the charge ordering have been observed~\cite{TSuzuki,Kimura,Kimura2}.
Notwithstanding no signal of the charge ordering by scattering technique, the magnetic order affects the character of the superconductivity in LSCO around $x=0.12$~\cite{Tsuzuki2}.
Therefore, a change in electronic states must appear more or less in case of the orthorhombic phase.
It is necessary for full-understanding of this problem  to discuss changes in the electronic state and in the crystal structure by investigating the detailed behavior of the transport properties of so-called stripe phase in the TLT phase and of the magnetically ordered state in the OMT phase.
More recently, Noda {\it et al}. reported the temperature dependence of the Hall coefficient in the stripe phase La$_{2-x-y}$Nd$_{y}$Sr$_x$CuO$_4$~\cite{Noda}.
However, the detailed behavior of the Hall coefficient at low temperature below 25 K is unknown yet.
In this paper, we report the experimental results of the temperature dependence  of the Hall coefficient at low temperatures and under high magnetic fields in the TLT and in the OMT phase.
\\
\indent
Samples used in this work, La$_{1.84}$Y$_{0.04}$Sr$_{0.12}$CuO$_4$ (LYSCO) and La$_{1.88}$Sr$_{0.12}$CuO$_4$ (LSCO) , are single crystals grown by the traveling-solvent-floating-zone (TSFZ) method under oxygen atmosphere.
The resistivity and the Hall coefficient are measured by usual four-probe dc method.
Six probes are putted on one sample and the Hall coefficient and the resistivity are measured simultaneously.
Current and magnetic field are reversed in each temperature to eliminate thermoelectric voltage of lines and resistive voltage between Hall probes.
The typical size of measured samples is $\sim 0.25\times 1\times 3$ mm.
The direction of the current (20 mA) is parallel to the CuO$_2$ plane and magnetic fields are applied perpendicular to the CuO$_2$ plane.
Neutron scattering measurement in LYSCO is carried out using the KSD double-axis spectrometer installed in the JRR-3M Guide Hall at the JAERI in Tokai, Japan.
The incident neutron beam has a wavelength of 1.53 \AA, obtained using a PG(002) monochromator.
The horizontal divergence of incident neutron beam is 12$^{\prime}$ and the acceptance angle of scattered beam is 30$^{\prime}$.
The sound velocity {\it V}$_{\rm s}$ is measured by the phase comparison method with the $\sim$12 MHz longitudinal waves generated by the PZT transducer.
All measurements of the temperature dependence under magnetic fields in this study were carried out with increasing temperature after field-cooling from 80 K in each time.
\\
\indent
The dc magnetization measurement is carried out in the two samples by a SQUID magnetometer.
Sharp superconducting transition is observed in both compounds, and the transition temperature {\it T}$_{\rm c}$ is 17 K in LYSCO and 27 K in LSCO, respectively.
\\
\indent
The upper panel of Fig.1 shows the temperature dependence of the neutron scattering intensity at (300) position of La$_{1.84}$Y$_{0.04}$Sr$_{0.12}$CuO$_4$.
The index is defined as the notation in the tetragonal (I4/mmm).
The solid line is a guide for the eyes.
As decreasing temperature, the intensity begins to increases slightly at 90 K and shows the rapid increase at 65 K.
Comparing with results of the structural analysis in a similar rare-earth doped LSCO system~\cite{Buchner,Crawford}, two structural phase transition temperatures are determined as {\it T}$_{\rm LO}$=90 K and {\it T}$_{\rm LT}$=65 K in La$_{1.84}$Y$_{0.04}$Sr$_{0.12}$CuO$_4$.
{\it T}$_{\rm LO}$ is the structural phase transition temperature to the low-temperature orthorhombic (OLT) phase, and {\it T}$_{\rm LT}$ is that to the low-temperature tetragonal (TLT) phase.
\begin{figure}[hhh]
\includegraphics[40mm,25mm][55mm,115mm]{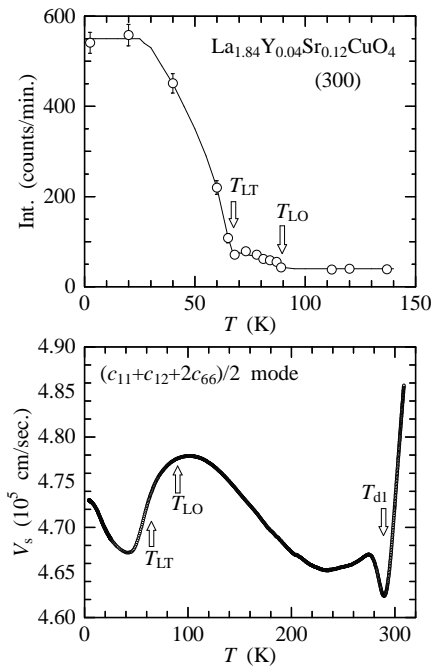}
\caption{\label{fig:epsart} Top : The temperature dependence of the neutron scattering intensity at (300) position in the tetragonal notation.
The solid line is guide for the eyes.
Bottom : The temperature dependence of the longitudinal sound velocity of La$_{1.84}$Y$_{0.04}$Sr$_{0.12}$CuO$_4$.
Open arrows in figures indicate structural phase transition temperatures.
The temperature {\it T}$_{\rm d1}$ is the structural phase transition temperature from the high-temperature tetragonal phase to the mid-temperature orthorhombic (OMT) phase.}
\end{figure}
The lower panel in Fig.1 shows the temperature dependence of the longitudinal sound velocity {\it V}$_{\rm s}$ corresponding to ({\it c}$_{11}$+{\it c}$_{12}$+2{\it c}$_{66}$)/2 mode in La$_{1.84}$Y$_{0.04}$Sr$_{0.12}$CuO$_4$.
On cooling, {\it V}$_{\rm s}$ begins to decrease at 90 K and increases below 40 K.
This type of change in {\it V}$_{\rm s}$ is the typical behavior around {\it T}$_{\rm LT}$~\cite{Suzuki1,Hiroshima}.
The large change of {\it V}$_{\rm s}$ suggests that the structural phase transition occurs in the bulk of this sample.
\\
\indent
Elastic magnetic incommensurate peaks are comfirmed at 2.1 K in LYSCO.
In the two samples which undergo the structural phase transition to the TLT phase, such as La$_{2-x-y}$Nd$_{y}$Sr$_{x}$CuO$_4$~\cite{Tranq1,Tranq2,Tranq3,Zimm} and La$_{1.875-x-y}$Ba$_{x}$Sr$_{y}$CuO$_4$~\cite{Goka}, the existence of magnetic and charge peaks are comfirmed by the elastic neutron scattering.
It is expected that the stripe phase appears in LYSCO at low temperatures~\cite{comment}.
\\
\indent
Figure 2 shows the temperature dependence of the Hall coefficient {\it R}$_{\rm H}$ of LYSCO and LSCO.
For comparison between the two compounds, {\it R}$_{\rm H}$ is normalized by the value at 100 K, which is 0.0055 cm$^3$/Coulomb in LYSCO and 0.0052 cm$^3$/Coulomb in LSCO.
The arrow in Fig.2 indicates the structural phase transition temperature {\it T}$_{\rm LT}$ of La$_{1.84}$Y$_{0.04}$Sr$_{0.12}$CuO$_4$.
As decreasing temperature in 4T, {\it R}$_{\rm H}$ of LYSCO shows the drastic decrease below the vicinity of {\it T}$_{\rm LT}$ and the sign reversal occurs in the low temperature region from 25 K to 15 K.
In the orthorhombic LSCO, only the gradual decrease in {\it R}$_{\rm H}$ is observed below 50 K which is far above {\it T}$_{\rm c}=$27 K, but no sign reversal is detected.
\begin{figure}[hhh]
\includegraphics[40mm,20mm][55mm,115mm]{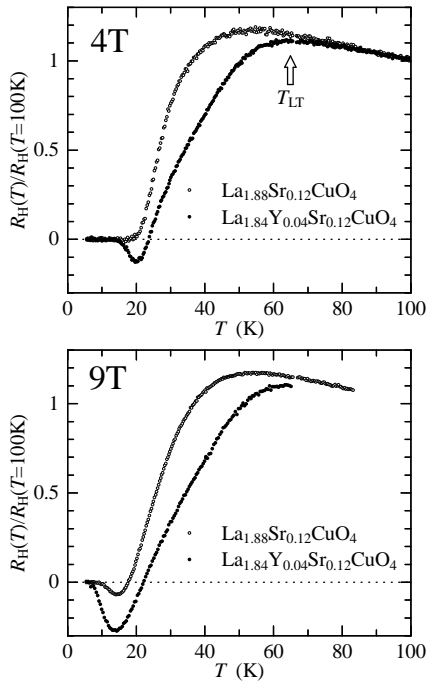}
\caption{\label{fig:epsart} The temperature dependence of the Hall coefficient of La$_{1.88-y}$Y$_y$Sr$_{0.12}$CuO$_4$ ($y=0$ : open circles, $y=0.04$ : closed circles) in 4 T and 9 T.
The arrow in the figure indicates the structural phase transition temperature to the TLT phase in the case of $y=0.04$.}
\end{figure}
These experimental results seem to be commonplace, since the anomalous change of {\it R}$_{\rm H}$ is observed in the TLT phase but not in the OMT phase and {\it R}$_{\rm H}$ in the TLT phase goes to zero in low temperatures as expected in the simplified stripe model~\cite{Noda}.
In 9 T, however, the sign reversal of {\it R}$_{\rm H}$ in low temperatures is enhanced.
In LYSCO, the temperature range where the sign is negative spreads and the absolute value at minimum point becomes larger, though the temperature where {\it R}$_{\rm H}$ begins to fall is not changed in 9 T.
Surprisingly in the orthorhombic LSCO, the sign reversal does appear below 20 K though no change of the sign is observed in 4 T.
\\
\indent
The discontinuous change of {\it R}$_{\rm H}$ at {\it T}$_{\rm LT}$, which is observed in La$_{2-x-y}$Nd$_{y}$Sr$_{x}$CuO$_4$ for $x<$1/8~\cite{Uchida,Noda}, does not appear in LYSCO.
It seems that the broadeness of the structural phase transition vignettes the discontinuity in {\it R}$_{\rm H}$ at {\it T}$_{\rm LT}$.
\\
\begin{figure}[hhh]
\includegraphics[40mm,30mm][55mm,95mm]{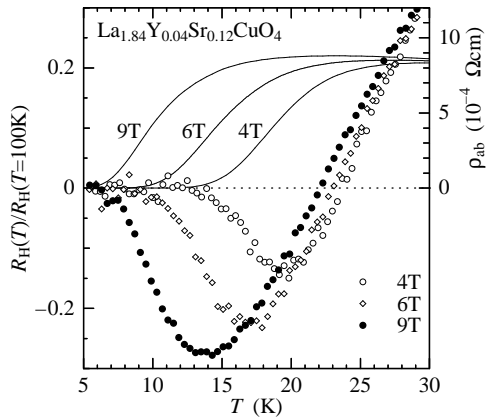}
\caption{\label{fig:epsart} The temperature dependence of the Hall coefficient and the resistivity of La$_{1.84}$Y$_{0.04}$Sr$_{0.12}$CuO$_4$ in 4 T, 6 T and 9 T.
Marks and solid lines denote the Hall coefficient and the resistivity, respectively.}
\end{figure}
\indent
Figure 3 shows the temperature dependence of {\it R}$_{\rm H}$ and the resistivity $\rho_{\rm ab}$ of LYSCO in 4 T, 6 T and 9 T.
The anomalous change of {\it R}$_{\rm H}$ becomes more significant as reducing the superconducting transition temperature by magnetic fields.
As decreasing temperature, {\it R}$_{\rm H}$ changes its sign and goes to zero along a curve similar to the temperature dependence of $\rho_{\rm ab}$.
Quantitatively, the temperature where d$\rho_{\rm ab}$/d{\it T} becomes maximum coincides with that of $-$d{\it R}$_{\rm H}$/d{\it T} in each magnetic field.
Therefore, the reason that {\it R}$_{\rm H}$ goes to zero from the negative side at low temperature is the superconductivity, and the real behavior of {\it R}$_{\rm H}$ is concealed by zero-resistivity.
We conclude that the Hall coefficient in the ground state of the stripe phase of La$_{1.84}$Y$_{0.04}$Sr$_{0.12}$CuO$_4$ has a finite negative value when the superconductivity is suppressed by magnetic fields.
Similar temperature and magnetic field dependence of {\it R}$_{\rm H}$ is reported in La$_{2-x}$Ba$_{x}$CuO$_4$ ($x=0.11$) single crystal which undergoes the structural phase transition to the TLT phase~\cite{Adachi}.
\\
\indent
Our measurement results are consistent with the new theoretical calculation by Prelov\v{s}ek {\it et al}.~\cite{Maekawa}.
They reported the numerical calculation within the {\it t}-{\it J} model, and conclude that {\it R}$_{\rm H}<$0 for $x<$1/8 and {\it R}$_{\rm H}\sim$0 for $x=$1/8 at {\it T}=0 in the stripe phase.
\\
\indent
Here, we touch upon the recent report on measurements of {\it R}$_{\rm H}$ in La$_{2-x-y}$Nd$_{y}$Sr$_{x}$CuO$_4$~\cite{Noda}.
They insist that {\it R}$_{\rm H}$ goes to zero at low temperature in the stripe phase of $x=0.12$ and that the charge transport is one-dimensional, but their measurement is stopped at 25 K at which {\it R}$_{\rm H}$ changes its sign in LYSCO(fig.3).
Measurements below 25 K under high magnetic fields should be carried out in order to discuss the behavior in the ground state of the stripe phase.
\\
\begin{figure}[hhh]
\includegraphics[40mm,30mm][55mm,95mm]{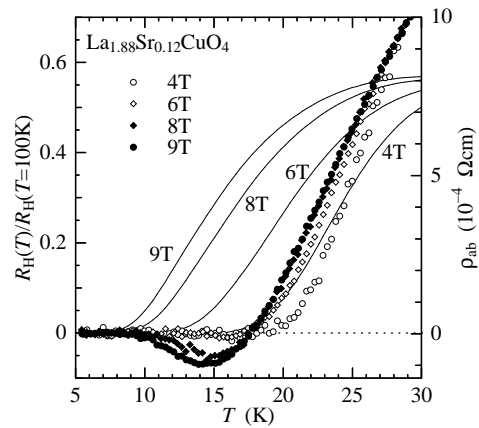}
\caption{\label{fig:epsart} The temperature dependence of the Hall coefficient and the resistivity  of La$_{1.88}$Sr$_{0.12}$CuO$_4$ in 4 T, 6 T, 8 T and 9 T.
Marks and solid lines denote the Hall coefficient and the resistivity, respectively.}
\end{figure}
\indent
Figure 4 shows the temperature dependence of the Hall coefficient and the resistivity of La$_{1.88}$Sr$_{0.12}$CuO$_4$ in various magnetic fields.
The sign reversal occurs in magnetic fields above 8 T.
By analogy with the discussion in the case of LYSCO, it is in consequence of the suppression of the superconductivity by magnetic fields. 
This result indicates that, also in the orthorhombic phase, {\it R}$_{\rm H}$ has a negative value at low temperatures in the hole concentration of $x\sim0.12$.\\
\indent
One thing which we would like to mention is that the origin of the negative {\it R}$_{\rm H}$ is by no means the contribution of vortex dynamics.
The reason is as follows:
(a) In high-{\it T}$_{\rm c}$ cuprates, the sign reversal of {\it R}$_{\rm H}$ in the vortex states tends to vanish as increasing magnetic fields~\cite{Matsuda1,Matsuda2}.
The data in this study is contrary to this tendency.
(b) As decreasing temperature, {\it R}$_{\rm H}$ begins to decrease at the vicinity of {\it T}$_{\rm LT}$ in LYSCO or at $\sim$50 K in LSCO rather than {\it T}$_{\rm c}$, and changes its sign continuously in each magnetic field.
(c) In LYSCO, the sign reversal of {\it R}$_{\rm H}$ occurs above {\it T}$_{\rm c}$ in zero-field.
(d) In policrystalline samples of LYSCO, this anomaly is significant around $x=0.115$~\cite{Geka}.
Therefore, we conclude that the sign reversal observed in this study is the intrinsic phenomena in the stripe phase.
\\
\indent
In the last, we discuss the temperature dependence of the thermoelectric power, which shows a similar temperature dependence~\cite{Sera,Uchida}.
From the fact that the anomalous change of the Hall coefficient and of the thermoelectric power is the most significant around the hole concentration of $x=0.12$, the origin of this anomalous change will be the same electronic state appeared below {\it T}$_{\rm LT}$.
Therefore, the thermoelectric power will shows the enhancement of this anomaly in high magnetic fields.
\\
\indent
In summary, the Hall coefficient of La$_{1.84}$Y$_{0.04}$Sr$_{0.12}$CuO$_4$ and La$_{1.88}$Sr$_{0.12}$CuO$_4$ is measured in high magnetic fields up to 9 T.
In the low-temperature tetragonal phase, as decreasing temperature, the Hall coefficient decreases and changes its sign at low temperatures.
In the orthorhombic La$_{1.88}$Sr$_{0.12}$CuO$_4$, the sign reversal of the Hall coefficient appears under high magnetic fields above 8 T.
The anomalous behavior of the Hall coefficient, that is the sign reversal, is enhanced in the condition that the superconductivity is suppressed by magnetic fields applied to the direction perpendicular to the CuO$_2$ plane.
Therefore, the Hall coefficient has a finite negative value in the ground state of the stripe phase.
\\
\indent
The Hall coefficient and the resistivity in high magnetic fields were carried out at the Center for Low Temperature Science, Tohoku University.
The authors are grateful to T. Nojima for measurement using superconducting magnet.
This research was partially supported by the Minstry of Education, Science, 
Sports and Culture, Grant-in-Aid for Scientific Research on Priority Areas 
(Novel Quantum Phenomena in Transition Metal Oxides), 12046256, 2000,
and for Scientific Research (C), 12040360, 2000.


\begin{references}
\bibitem{comment} The anomalous change of the Hall coefficient is more significant in LYSCO than in LSCO.
Therefore, the whole argument in this report is not affected even if there exist a residual orthorhombic (Pccn or OMT) phase at low temperature in LYSCO.
\end{references}
\end{document}